\documentclass[english]{article}

\usepackage[utf8]{inputenc}
\usepackage[T1]{fontenc}
\usepackage{babel}

\usepackage[doi=true, isbn=false, url=false, eprint=false, maxbibnames=99]{biblatex}
\usepackage{booktabs}
\usepackage{tablefootnote}
\usepackage{nicematrix}
\usepackage[pdfborder={0 0 0}]{hyperref}
\usepackage{cleveref}

\usepackage[a4paper, left=1in, right=1in, top=1in, bottom=1in, includeheadfoot, showframe=false]{geometry}

\usepackage{enumitem}
\usepackage[format=plain, labelfont={bf}, textfont=bf]{caption}

\usepackage{fancyhdr}
\fancypagestyle{plain}{%
\fancyhf{}
\fancyhead[R]{\thepage} 
\lhead{Journal of Open Research Software}
}
\usepackage{parskip}

\usepackage{tikz}
\usetikzlibrary{matrix, positioning, patterns}

\date{}

\usepackage{charter}

\usepackage{siunitx}

\usepackage{cleveref}
\usepackage{booktabs}

\usepackage{tikz}
\usetikzlibrary{automata, positioning, arrows, arrows.meta, decorations.pathreplacing, calc, chains, shapes.misc, backgrounds, fit, perspective, matrix, shapes, shadows.blur, shapes.multipart, patterns, decorations.pathmorphing, 3d, fadings}

\usepackage[tikz]{mdframed}

\title{No Free Lunch: Research Software Testing in Teaching}
\author{%
Michael Dorner, Andreas Bauer, and Florian Angermeir
} 

\begin{document}

\maketitle

\begin{abstract}
Software is at the core of most scientific discoveries today. Therefore, the quality of research results highly depends on the quality of the research software. Rigorous testing, as we know it from software engineering in the industry, could ensure the quality of the research software but it also requires a substantial effort that is often not rewarded in academia.
%
Therefore, this research explores the effects of research software testing integrated into teaching on research software.
%
In an \emph{in-vivo} experiment, we integrated the engineering of a test suite for a large-scale network simulation as group projects into a course on software testing at the Blekinge Institute of Technology, Sweden, and qualitatively measured the effects of this integration on the research software. 
%
We found that the research software benefited from the integration through substantially improved documentation and fewer hardware and software dependencies. However, this integration was effortful and although the student teams developed elegant and thoughtful test suites, no code by students went directly into the research software since we were not able to make the integration back into the research software obligatory or even remunerative.
%
We strongly believe that integrating research software engineering, such as testing, into teaching is valuable not only for the research software itself but also for students gaining exposure to bleeding-edge research in their field. However, the uncertainty about the intellectual properties of students' code contributions substantially limits the potential of integrating research software engineering into teaching. %
\end{abstract}

\section{Introduction}

Software underpins almost all research and is, in one way or another, at the core of most scientific discoveries today. Researchers from different disciplines develop new or customize existing software to collect, analyze, and visualize data to push the limits of the unknown. Thus, the quality of research results highly depends on the quality of the research software. 

Although the significance of software testing for quality assurance is widely acknowledged \cite{garousi2016SystematicLiteratureReview, garousi2017WhatIndustryWants, sanchez-gordon2018CertificationsInternationalStandards}, efforts to ensure the quality of research software are often not rewarded in academia~\cite{Eisty2022} and are rarely included in researchers' training~\cite{Nangia2017}.

To address this major shortcoming of research software engineering, we integrated research software testing into teaching aiming for synergy effects between teaching and research software testing for improving the quality and trustworthiness of real-world research software. In this article, we report the effects of such an integration on research software. In detail, we set out to answer the following research question: \\

\begin{mdframed}[userdefinedwidth=\textwidth, align=center]
RQ: What are the effects of integrating research software testing into teaching on research software? 
\end{mdframed}

To answer our research question, we conducted an \emph{in-vivo} experiment on such an integration. In the context of a course on software testing at Blekinge Institute of Technology, Sweden, students developed a test suite for real-world research software. We gather and analyze qualitative observations on the effects of the integration on the research software. 

The remainder of this article is structured as follows: In \Cref{sec:background}, we briefly discuss the background of our study. We then describe our experimental design with its core components in \Cref{sec:experimentaldesign} in detail. In \Cref{sec:results}, we report the resulting effects of our experiment on the research software (\Cref{sec:results}). After discussing the limitations of our study (\Cref{sec:limitations}), we conclude our paper by raising a fundamental question in the context of integrating research software engineering into teaching: How to handle the intellectual property of students properly?

\section{Background}
\label{sec:background}

Research on software-testing education has gained more attention over the last years \cite{Garousi2020}. In light of our experiment on integrating research software testing into teaching, we would like to highlight the current approaches and initiatives to foster research software engineering in teaching (\Cref{sec:researchsoftwareengineeringinteaching}) and using open-source software in software-testing education (\Cref{sec:ossforsoftwaretestingeducation}).

\subsection{Research Software Engineering in Teaching}
\label{sec:researchsoftwareengineeringinteaching}

As the research software engineering publication monitor shows, research software engineering has gained more attention over the recent years \cite{fritz2024}. A proper education has become one of the key concerns in research software engineering \cite{cohen2021}. This need for education driven by different organizations such as \emph{HIFIS}\footnote{\url{https://hifis.net/services/software/training.html}}, \emph{INTERSECT}\footnote{\url{https://intersect-training.org/}}, \emph{CodeRefinery}\footnote{\url{https://coderefinery.org/}}, or \emph{BSSw}\footnote{\url{https://bssw.io/}} address this concern by providing learning materials, offering workshops or teaching, and maintaining communities to share research software engineering teaching experiences. 

Despite its importance for modern science, surprisingly, research software engineering is rarely a part of curricula. General courses on computer science education (e.g., \emph{The Missing Semester of Your CS Education}\footnote{\url{https://missing.csail.mit.edu/}} at MIT or the \emph{The Missing CS Class} at University of California, Davis, \cite{gilson2022}) or courses specific to research software engineering (e.g., University of Potsdam \cite{potsdam2024}, Technical University of Dresden \cite{dresden2024} are notable and well-received exceptions.

Even though there is progress towards addressing education, we agree with \citeauthor{goth2024} that a systemic institutionalization of research software engineering education is needed \cite{goth2024}.

\subsection{Open-Source Software for Teaching Software Testing}
\label{sec:ossforsoftwaretestingeducation}

Providing authentic, real-world software quality assurance and testing experiences within the context of a Computer Science or Software Engineering curriculum is challenging \cite{Chen2014, Deng2020}. Different studies used open-source projects for this reason \cite{Chen2014, Krutz2014, Deng2020, Venson2024}. Open-source software is software that is available under a license that grants the right to use, modify, and distribute the software---modified or not---to everyone free of charge (adapted from \cite{Riehle2023}).

None of the studies has addressed the ethical and potentially legal issue we will discover in the course of this article: Contributing to open-source projects requires the copyright of students, which puts us researchers and teachers in a conflicting position since we (teachers) grade students for work that we (researchers) exploit.  

\citeauthor{Venson2024} proposed and discussed their teaching approach for a software testing course that integrates theory and practical experience through the utilization of both team-based learning and active contributions to open-source software projects \cite{Venson2024}. Although not in the context of research software testing, we can partially confirm their reflections and recommendations in our experiment: Almost all students had no experience with contributing to open-source software. Therefore, the quality of documentation plays a pivotal role in the integration of open-source software into testing courses. 

In contrast to our experiment, \citeauthor{Venson2024}~\cite{Venson2024} awarded students for code contributions via pull requests with extra points. Pull requests are a convenient way for software developers to collaborate on open-source projects by proposing changes\footnote{\url{https://docs.github.com/en/pull-requests}}. However, they observed that \textit{``this incentive prompted students to submit pull requests lacking genuine contributions, which inundated the [open-source software] community with superfluous requests''}. As a result, they discontinued offering extra points for pull requests. In our work, we did not incentivize contributions, for example by offering bonus points or other rewards, and could not observe any code contribution in the form of a pull request.


\section{Experimental Design}
\label{sec:experimentaldesign}

In this section, we describe the design of our \emph{in-vivo} experiment that aims to answer our research questions on the effects of integrating research software engineering into teaching. 

An \emph{in-vivo} experiment (or field experiment) is an experiment conducted in a natural setting with a high degree of realism. The researcher manipulates some properties or variables in the research setting to observe an effect of some kind. The realistic research setting exists independent of the researcher, which distinguishes it from the contrived research setting in a laboratory experiment \cite{Stol2018}. 

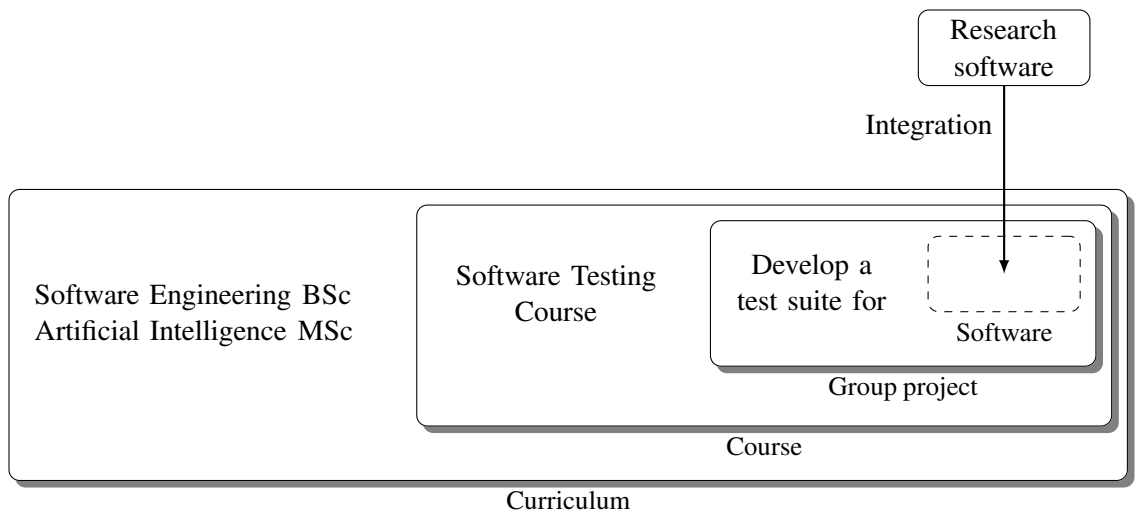
\begin{figure*}
\centering
\begin{tikzpicture}[%
	node distance = 1em and 1em,
]

\node[draw, dashed, rounded corners, minimum height=1cm, minimum width=2cm, inner sep=2mm, label={[font=\small, name=softwarelabel]below:Software}] (placeholder) at (0,0) {};

\node[base left= of placeholder, minimum height=1cm, text width=2cm, text centered] (task)  {Develop a test suite for};

\node[draw, rounded corners, fit=(task)(placeholder)(softwarelabel), inner sep=2mm, label={[font=\small, name=groupprojectlabel]below:Group project}] (groupproject) {};

\node[left=of groupproject, text width=3cm, text centered] (course) {Software~Testing\\Course};

\node[draw, rounded corners, fit=(groupproject)(course)(groupprojectlabel), inner sep=2mm, label={[font=\small, name=courselabel]below:Course}] (projectpluscourse) {};

\node[left= of projectpluscourse, minimum height=3em, text width=4.5cm] (curriculum) {Software Engineering BSc\\Artificial Intelligence MSc};

\node[draw, rounded corners, fit=(curriculum)(projectpluscourse)(courselabel), inner sep=2mm, label={[font=\small, name=curriculumlabel]below:Curriculum}] (curriculumpluscourse) {};

\node[draw, rounded corners, minimum height=1cm, text width=2cm, text centered] at (0,3) (simulation) {Research software};

\draw[thick, -latex] (simulation) to[out=-90, in=90,] node[pos=0.2, left] {Integration} (placeholder.center);

\begin{scope}[on background layer]
\node[rounded corners, fit=(curriculumpluscourse), inner sep=0mm, drop shadow, fill=white] {};
\node[rounded corners, fit=(projectpluscourse), inner sep=0mm, drop shadow, fill=white] {};
\node[rounded corners, fit=(groupproject), inner sep=0mm, drop shadow, fill=white] {};
\end{scope}

\end{tikzpicture}
\caption{Overview of the research setting of our experiment.}
\label{fig:experimentaldesign}
\end{figure*}

In our experiment, the natural setting of our integration is a course on software testing, which is part of the regular curriculum for the Software Engineering and Artificial Intelligence study programs at Blekinge Institute of Technology, Sweden. This course comprises a project work in which the students develop a test suite for software. Instead of a classical software project or an open-source component, as it is commonly used in teaching \cite{Chen2014, Krutz2014, Deng2020, Venson2024}, we used a research software for our experiment and observed the effects of this integration on this research software. In the context of an experiment, we consider our research software as a type of real-world example and not as a toy project or educational demonstrator that may have limitations in terms of generalizability to real-world issues~\cite{wohlin2012ExperimentationSoftwareEngineering}. Thus, we see research software as a special type of real-world open-source software. \Cref{fig:experimentaldesign} provides an overview of the setting of our experiment.

In the following sections, we describe the two core components of our experimental design, the setting of our experiment and the subject, the research software.

\subsection{Setting}
\label{sec:setting}

The setting for our experiment is a software testing course for students of two study programs at Blekinge Institute of Technology: Bachelor of Science (BSc) in Software Engineering and Master of Science (MSc) in Artificial Intelligence. The course was mandatory for students in the Bachelor's program in Software Engineering but optional for those enrolled in the Master's program in Artificial Intelligence.

This software testing course introduces hands-on testing and quality assurance techniques for software systems. The course aims to make its participants realize how testing can improve software quality if it is effectively integrated into the software development processes and understand how this can be achieved using both established and new techniques in software testing. The aim is also to convey practical experience with tools that support and automate these techniques.

As such, the course is split into two parts. In the first part, ten lectures provide the theoretical foundations of software testing. Two practical assignments deepen the understanding of the concepts introduced in the lectures and introduce students to the technical aspects of software testing in Python. In the second part of the course, students work in groups and develop a test suite for a software project. In previous course iterations, the software projects ranged from open-source to projects that students developed. Afterward, the students summarize their findings as a written report.

We conducted the experiment in the Spring of 2023 with \num{40} enrolled students. Twelve teams worked in three to four student groups on the given tasks.

\subsection{Research Software} 
\label{sec:researchsoftware}

As part of the experiment, we asked the students to develop a test suite for research software that we developed previously to simulate information diffusion in code reviews at Microsoft, Spotify, and Trivago modeled as communication networks\footnote{\url{https://github.com/michaeldorner/information-diffusion-boundaries-in-code-review}}~\cite{Dorner2023software}. 

The research software models the communication networks emerging from code review as time-varying hypergraphs in which each vertex represents a developer and each edge represents a code review, which can connect multiple developers exchanging information. Hypergraphs are a generalization of traditional graphs and capture time-dependent and higher-order interactions in social and communication networks in which edges may connect more than two vertices. 

The time dependency also extends the notion of a minimal distance in time-varying hypergraphs: a path cannot only be the shortest in terms of its topological distance but also the fastest or foremost path in terms of the temporal distance. Dijkstra’s algorithm can be used to find the minimal paths and, therefore, minimal distances in such graphs. However, Dijkstra's algorithm for time-varying hypergraphs has not been described in the literature or implemented before. 

The communication networks modeled as time-varying hypergraphs in our simulation can become huge. For example, the communication network emerging from Microsoft's code review consists of over \num{309740} code reviews (edges) and \num{37103} developers (vertices). The simulation is, therefore, computationally expensive: On conventional hardware, it takes several days to simulate the information diffusion and to process the results. 

Before we started the experiment, the simulation had a minimal testing suite covering tests for the graph data structure but not for the algorithm. Furthermore, a minimal but complete \texttt{README} file contained instructions on the installation and how to use it. As the course started, the research software was already publicly available, but the related publication \cite{Dorner2023} was not published. 

The simulation fits well for our experiment for the following reasons: 

\begin{itemize}
\item The minimal existing test suite helped the students to get started without revealing too much of the required solution. Yet, there is no single solution to the problems, but different solution paths. 
\item The algorithm is novel. The lack of a reference implementation makes thorough testing non-trivial and yet very important. 
\item Manual testing does not scale with the sheer size and complexity of the communication networks. Also, brute-force testing approaches are not feasible due to the computational complexity.
\end{itemize}

We defined requirements for the students to implement the test suite for the research project as follows:

\begin{itemize}
\item The test suite must comprise unit tests to automatically test the research software.
\item Test coverage metrics must be reported and discussed.
\item The test suite must cover two Python version (3.8 and 3.11).
\item Beyond these minimal requirements, each team must select a focus of their overall test suite. Possible areas to focus on are performance testing, unit and integration testing, fuzzing, or visual testing. 
\end{itemize}

Before the group-project phase of the course started, we informed the students about the experiment and provided those instructions along with the \texttt{README} and the scientific literature that introduced the mathematical background \cite{Dorner2022, Dorner2023} to the student teams. After the first week, the student teams decided as a team for one focus area of the test suite.

\section{Results}
\label{sec:results}

We observed three effects on our research software as a result of integrating software testing into teaching. Although these effects—such as improved documentation and reduced hardware and software dependencies—might have occurred over time, they were specifically triggered by the course integration. Therefore, they are direct outcomes of this integration and are detailed in the following sections.

\subsection{Tests but No Test Code Integration} 

The student teams tested the research software extensively and developed diverse and well-engineered testing suites for our research software as part of their projects. The implemented unit and integration tests validate the expected behavior of the research software. Two project groups extended the suite beyond classical testing to catch memory corruption and safety bugs through fuzzy testing. Although no bugs were found, the different testing approaches and diverse implementations of test suits provide a comprehensive perspective on the quality of the research software. As authors of the research software, we would not have been able to provide such in-depth testing for the research software.

However, we could not integrate any test code developed by students into the main project. This shortcoming is rooted in the way contributions to open-source projects are handled. Our research software is open-source and licensed under the the so-called MIT license\footnote{\url{https://opensource.org/license/mit}}, aligning with best practices for research software development~\cite{Mendez2020}. Contributing to any open-source project requires the contributor, as the owner of the intellectual property, to grant permission, per the project's open-source license, to use, copy, distribute, or modify their contributions free of charge. That implies, researchers are only allowed to use student contributions if the students make their code available under the same or a compatible open-source license. In our academic setting, an inherent conflict arises: Teachers evaluate the students' contributions while also benefiting from their contributions. Thus, the teachers (as research software authors) exploit those contributions, although not necessarily for financial gain. Hence, an ethical conflict arises since students may feel forced to grant permission for their intellectual property to avoid any potential negative impact on their grading. In jurisdictions like Germany or Sweden, this also becomes a legal issue. To avoid ethical and legal issues, we reached out for legal advice and to the student council, seeking support. All parties advised us to underline the optional and voluntary character of the potential contributions to the research software.

Another obstacle for group projects is that all group members must agree to publish their code under an open-source license. It would be nearly impossible to unravel contributions from individual students without at least touching the intellectual property of other team members. We did not explore how willing individual group members would have been to contribute to the main project while avoiding potential threats to the voluntary nature of contributions.

Additionally, integrating code into the main software project is not effortless. Very rarely, code is flawless and can be merged into the main software project without further considerations, changes, or discussions. At this point, we would like to emphasize that we feel sympathetic towards the students: Spending additional and yet significant efforts that are not rewarded and might be acknowledged in an academic context as they would become co-authors of research software is not very compelling. 

In summary, our research software was tested thoroughly in this course, although we could not motivate the student groups to integrate their code into the main research software.

\subsection{Reduced Hardware and Software Dependencies}

We did not provide a unified hardware setup but relied on the students' computers for testing the research software. To cover the large diversity in hardware, operating systems, and software environments, we removed all operating system and minimized hardware requirements and code dependencies. 

Before the course, the research software was tested on macOS and Linux (Ubuntu). In preparation for the course, we tested the research software on the common operating systems Windows 11, macOS, and Linux (Ubuntu) to avoid complications with students' operating systems and development environments. 

We also removed one software component and made it optional since. Although the component \texttt{orjson}\footnote{\url{https://github.com/ijl/orjson}} improves the performance of the file processing substantially,  \texttt{json}\footnote{\url{https://docs.python.org/3/library/json.html}} from the Python standard library serves the same purpose and requires no additional steps to use. 

Being agnostic on hardware and operating systems and having fewer external dependencies contribute to easier use, maintenance, and extension of our research software, facilitating a broader adoption among the research community. Further, minimizing software dependencies can decrease the fault proneness of a software system~\cite{Cataldo2009}. 

For us as developers of research software, minimizing hardware and software created a small additional effort that is worthwhile to increase the re-usability of research software.

\subsection{Improved Documentation}

We found that scientific literature \cite{Dorner2022, Dorner2023} and a minimal documentation in the form of a \texttt{README} file were not sufficient for its use in the context of teaching. To ease the barriers for students at the project start and during the course, we continuously improved the documentation in size and quality. We observed that continuously improving documentation enabled efficient communication with the students. 

Although the documentation improved, less experienced students requested additional support to understand the domain and technical details of research software, which is crucial for an efficient test suite. Therefore, we offered each group weekly and individual students one-on-one sessions on request. In particular, during the first weeks, this offer was well received, and all groups booked meeting slots. 

The continuous support through additional weekly meetings for all groups and the continuous improvement of the documentation required significant additional efforts, which exceeded the efforts that would have been required for an open-source project, for example, where the domain and functionality are more obvious. 

We believe the improved documentation will help researchers reproduce or replicate our simulation, ultimately fostering scientific growth through traceability, transparency, and usability. This finding confirms the reflection and recommendation by \citeauthor{Venson2024} \cite{Venson2024}.

\section{Limitations}
\label{sec:limitations}

As for any \emph{in-vivo} experiment, there are two inherent limitations \cite{Stol2018}: First, our findings are limited in their transferability to other settings, which means other courses and also other research software projects. Second, although the setting of the experiment is realistic, it is subject to confounding factors such as the specific type of research software or previously attended courses that limit the precision of measurement. However, we believe that our high-level findings are still transferrable to other settings since the course introduces the students to the foundations of software testing and does not require prior knowledge of software testing. Additionally,  two different curricula participated: Bachelor of Science (BSc) in Software Engineering and Master of Science (MSc) in Artificial Intelligence, which broadens the transferability to other curricula. Last, the research software uses Python, including Jupyter Notebooks, which are common in research software and teaching.

A considerable limitation of this study is that we do not quantify the (additional) effort for the integration since we did not track the hours spent on the integration or previous courses. Our estimates are based on personal experiences and relative to previous instances of the courses. However, we believe that a relative effort, as reported in this study, also provides valuable insights for other research software developers and teachers, even though a more precise effect size is missing. 

We missed the chance to tailor the course evaluation after the course was completed to our experiment and, therefore, to get a more profound insight into the experiment from a student perspective. We plan to replicate our experiment for the next iterations of the course and then collect student feedback to better understand the challenges and benefits for students as future researchers and future research software engineers. 

Since our research software---like a large subset of research software in general---is licensed under an open-source license and is, therefore, open-source software, we assume an overlap in integrating open-source software into teaching environments. However, we focus exclusively on research software and its ethical implications without delving into this overlap since the software not only serves an educational purpose but also supports the broader goal of enhancing the quality and impact of research software.

\section{Conclusion}

In our experiment, we integrated the testing of research software into teaching and observed the effects on the research software. 

On the one hand, integrating research software engineering into teaching substantially improved the research software and its usage and trustworthiness. Some student teams developed outstanding and thoughtful test suites to ensure the quality of the research software. The scale and extent of testing would not have been possible within the research team that initially developed the software. 

Additionally, through the integration, we prepared the research software for researchers who would like to reuse or build upon our research software in the future. We improved the software documentation in preparation for the course to ease the students' onboarding. We also lowered hardware and software dependencies, which made the initial setup easier, reduced the complexity of the software, and, thereby, improved the maintainability of the software in the future. This makes the integration into teaching an excellent incubator for open science for research software engineering efforts. 

On the other hand, those positive effects did not come for free. Due to the complexity of the topic, we had to invest a significant amount of time in preparing the research software for the course and in providing support to student teams.  

The most significant challenge we observed is the uncertainty about the intellectual property of student code contributions to the project. Contributing to open-source projects requires copyright, which puts us researchers and teachers in a conflicting position: We (teachers) grade students for work that we (researchers) exploit---although maybe not financially. This conflict raises at least ethical, but also---in jurisdictions like Germany or Sweden---legal questions. We could not incentivize or motivate students to go the extra mile to integrate their excellent work into the research software. It remains unclear how this dilemma could be resolved. This uncertainty regarding students' contributions extends beyond research software and is observable in contributions through bachelor's and master's theses. 

To build on the findings of this study, future research should focus on two key directions: First, exploring the learning experiences of students is crucial, as this important aspect was not covered in our current work. This exploration can provide valuable insights into the educational impact of integrating research software testing into teaching. Second, interdisciplinary research is needed to examine the legal and ethical perspectives related to the issues we raised. Addressing these perspectives can help remove the existing uncertainty and guide more informed practices in this area.

Although there are not yet well-worn paths for integrating research software testing into teaching, we believe that testing research software and teaching can cross-fertilize. On the one hand, the research software gets an incubator before being reused or extended in the scientific community. On the other hand, students get in touch with the latest research in their respective fields very early on as part of their curriculum. This hands-on experience with research software engineering can help to train and prepare students, the next generation of researchers.

\section*{Acknowledgement}
We thank the participants of the deRSE'24 poster session for their insightful discussions. 
This work was supported by the KKS Foundation through the SERT Project (Research Profile Grant 2018/010) at Blekinge Institute of Technology.

\printbibliography

\end{document}